

\magnification=\magstep1
\openup 1\jot
\def\Gria{{1}}
\def\Omn{{2}}
\def\GMH{{3}}
\def\HQC{{4}}
\def\Grib{{5}}
\def\RH{{6}}
\def\SW{{7}}
\def\HarLH{{8}}
\def\HH{{9}}
\def\ADM{{10}}
\def\Pena{{11}}
\def\Penb{{12}}
\def\Ash{{13}}
\def\GMHB{{14}}

\rightline {UCSBTH-94-41}
\rightline {LA-UR-94-2101}
\rightline {CGPG-94/10-1}
\rightline {grqc/9410006}
\vskip 1truecm

\centerline{\bf Conservation Laws }
\centerline{\bf in the Quantum Mechanics of Closed Systems}
\vskip 0.5truecm

\centerline{James B. Hartle\footnote{$^*$}{hartle@cosmic.physics.ucsb}}
\vskip 2mm

\centerline{ \it Department of Physics, University of California}
\centerline{ \it Santa Barbara, CA 93106}

\centerline{ \it Theoretical Astrophysics, T-6, MSB288,
Los Alamos National Laboratory}
\centerline{ \it  Los Alamos,  NM 87545}

\centerline{\it Isaac
Newton Institute for the Mathematical Sciences, University of Cambridge}
\centerline{\it 20 Clarkson Road, Cambridge, U.K. CB3 0EH}

\vskip 3mm

\centerline{ Raymond Laflamme\footnote{$^{\dagger}$}
{laf@t6-serv.lanl.gov}}
\vskip 2mm

\centerline{ \it Theoretical Astrophysics, T-6, MSB288,
Los Alamos National Laboratory}
\centerline{ \it  Los Alamos,  NM 87545}

\centerline{\it Isaac Newton Institute for the Mathematical Sciences,
University of Cambridge}
\centerline{\it 20 Clarkson Road, Cambridge, U.K. CB3 0EH}

\vskip 3mm

\centerline{Donald Marolf\footnote{$^{\ddagger}$}
{marolf@cosmic.physics.ucsb.edu}}
\vskip 2mm

\centerline{ \it Center for Gravitational Physics
and Geometry, Physics Department,}
\centerline{ \it The Pennsylvania State
University, University Park, Pennsylvania 16802}

\vskip 1.0cm
\centerline{Abstract}
\vskip 3mm

{ \leftskip 10truemm \rightskip 10truemm
\openup -1\jot

We investigate conservation laws in the quantum mechanics of closed
systems.  We review an argument showing that exact decoherence implies
the exact conservation of quantities that commute with the Hamiltonian
including the total energy and total electric charge.  However, we also
show that decoherence severely limits the alternatives which can be
included in sets of histories which assess the conservation of these
quantities when they are not coupled to a long-range
field arising from a fundamental symmetry principle.
We then examine the realistic cases of electric charge coupled to the
electromagnetic field and mass coupled to spacetime curvature and show
that when alternative values of charge and mass decohere, they
always decohere exactly and are
exactly conserved as a consequence of their couplings to long-range
fields. Further, while decohering histories that describe fluctuations
in total charge and mass are also subject to the limitations mentioned
above, we show that these do not, in fact, restrict {\it physical}
alternatives and are therefore not really limitations at all.

\openup 1\jot}
\vskip 1 truecm
\noindent {\it PACS number(s):}03.65.Bz, 03.65.Ca, 03.65.Db, 98.80.Bp.
 \vfill\eject

\proclaim \uppercase\expandafter{\romannumeral1} Introduction.

Energy is conserved during the unitary evolution
$$
|\psi (t) \rangle = {\rm e}^{-iht/\hbar} |\psi(0)\rangle
\eqno (1.1)
$$
of a quantum state of an isolated subsystem of the universe
because the Hamiltonian, $h$, of that subsystem commutes with
the unitary time
evolution operator.  However, energy is not generally conserved
by the ``second law'' of quantum evolution that describes how
the state of a subsystem evolves when an ``ideal''
measurement of it is carried out.
If $|\psi\rangle$ is the state before an ideal
measurement, the state afterwards is ``reduced'' to
$$
|\psi\rangle \rightarrow
 {s_\alpha |\psi\rangle \over \| s_\alpha |\psi\rangle \|}
\eqno (1.2)
$$
where $s_\alpha$ is a Schr\"odinger-picture projection operator
corresponding to the measurement outcome -- one of a set of such
operators $\{ s_\alpha\}, \alpha= 1,2,\dots$ describing different
possible outcomes.  Even if $|\psi\rangle$ is an energy eigenstate,
the reduced state vector will generally not be an energy eigenstate except
in the special case that $s_\alpha$ commutes with the Hamiltonian
of the subsystem $h$.

More generally, if a sequence of measurements is carried out on the
subsystem at
times $t_1,...,t_n$, with outcomes represented by Heisenberg picture
projection operators $\{ s^k_{\alpha_k}(t_k) \}, k=1,...,n$, the joint
probability of a particular sequence of outcomes $\alpha\equiv
(\alpha_1,...,\alpha_n)$ is
$$
p(\alpha_1,\dots,\alpha_n) =
       \| s^n_{\alpha_n}(t_n) \dots s^1_{\alpha_1}(t_1) |\psi\rangle \|^2 .
\eqno (1.3)
$$
Energy is conserved if the joint probability vanishes whenever
measurements of the energy at two times disagree.
However, if measurements of quantities that do not commute with $h$
intervene between the two determinations of the energy, then that
joint probability will not be zero.  More precisely,
if $\{ s^h_{\alpha_l}(t_l) \}$ and  $\{ s^h_{\alpha_m}(t_m) \}$
project to the same set of ranges of $h$-eigenvalues,
the joint probability has nonzero entries for $\alpha_l \neq
\alpha_m$ when the intervening projections
do not commute with $h$. Energy is thus not necessarily
conserved in a sequence of measurements.

The conservation of energy by the unitary evolution and the  general
non-conservation by the reduction of the state vector are not
surprising.  The unitary law (1.1) describes the evolution of a
subsystem in isolation.  The reduction law (1.2) describes the evolution of
a subsystem interacting with another system which is measuring it.
Energy may be exchanged
between the measuring apparatus and the measured subsystem.

The familiar ``Copenhagen'' quantum mechanics of measured subsystems
sketched above is an approximation to a more general quantum
mechanics of closed systems [\Gria,\Omn,\GMH],
most  generally the universe as a whole.
It is an approximation that is appropriate when certain approximate
features of realistic measurement situations may be idealized as
exact\footnote{$^*$}{For example, as in [\HQC], Section II.10.}.
The most general predictions of quantum mechanics
are  the probabilities for individual members
of sets of alternative histories of a closed system.  One kind
of alternative history set may be specified by giving exhaustive
sets of exclusive alternatives at a sequence of times $t_1,\dots,t_n$.
The alternatives at one time are represented by a set of
Heisenberg-picture projection
operators $\{P^k_{\alpha_k} (t_k) \}$ satisfying
$$
P^k_{\alpha_k} (t_k)P^{k}_{\alpha'{}_{k}} (t_k)=\delta_{\alpha_k
\alpha'{}_k}P^k_{\alpha_k}(t_k), \ \
\sum_{\alpha_k} P^k_{\alpha_k} (t_k) = I
\eqno (1.4)
$$
showing that they represent a mutually exclusive, exhaustive set of
alternatives.   An individual history corresponds to a particular
sequence of alternatives $\alpha\equiv (\alpha_1,\dots,\alpha_n)$ and is
represented by the corresponding chain of projections.
When the theory assigns probabilities to a set of such alternative
histories, the probabilities of the individual histories are given by
$$
p(\alpha_1,\dots,\alpha_n)=
 \|P^n_{\alpha_n}(t_n) \dots P^1_{\alpha_1}(t_1) |\Psi\rangle \|^2
\eqno (1.5)
$$
assuming (for simplicity) that the initial condition of the closed system
$|\Psi\rangle$ is  pure.

Eq.(1.5), giving the probability of a history of a closed system,
has the same form as eq.(1.2) giving the probability of a history of
measurements of a subsystem.  The only difference between the
expressions is that in
(1.2) states operators, {\it etc}, all act on the Hilbert space of the
measured subsystem, while in (1.5) they act on the Hilbert space
of a closed system  including  both the measured subsystem and any
measurement apparatus. Thus, if
$\{P^H_{\alpha_l} (t_l)\}$ and  $\{P^H_{\alpha_m} (t_m)\}$
are projectors onto the same sets of ranges of the total
energy $H$ at two different
times there is no reason to believe that the probability of histories
with $\alpha_l\neq\alpha_m$ will vanish if the intervening projections
do not commute with the Hamiltonian.  Eq.(1.5) no more conserves energy
than does (1.2).  However, in the quantum mechanics
of a closed  system there is nothing ``external'' to cause a
fluctuation in the total energy.  Does this mean that the
quantum mechanics of closed systems predicts non-zero probabilities
for violations of energy conservation?  Further,  not only
conservation of energy is at stake.  Similar remarks
hold for any other quantity that commutes with the Hamiltonian such
as the total electric charge\footnote{$^*$}{Following the usual
terminology, when no confusion should result, we will often refer to
quantities that commute with the Hamiltonian as ``conserved quantities"
even though it is their conservation that
is being investigated!}.  In the following, we shall
show that no such violations are in fact predicted.

In posing the question of possible violations of fundamental
conservation laws in the quantum mechanics of closed systems we should
stress that we do not mean violations that
might be revealed by successive measurements of a {\it subsystem}.
The probabilities
of the outcomes of ideal measurements on a subsystem are described by
(1.3) to an accuracy far beyond the precision available in any experimental
check of a conservation law.  A sequence of two measurements that
determines whether the value of a quantity $a$ that commutes with $h$ is in one
of a set
of ranges $\{\Delta_\alpha\}$ is represented by the string of projections
$$
s^a_{\alpha_2}(t_2) s^a_{\alpha_1}(t_1).
\eqno (1.6)
$$
The Heisenberg equations of motion
$$
s^a_{\alpha}(t)= {\rm e}^{iht/\hbar}s^a_{\alpha}(0) {\rm e}^{-iht/\hbar}
\eqno (1.7)
$$
together with the analog of (1.4) show that, when $a$ commutes with the
Hamiltonian $h$, the operator string (1.6) is proportional
to $\delta_{\alpha_1 \alpha_2}$.  The probabilities (1.5)
of a measured fluctuation
in the value of a quantity commuting with $h$, including the energy itself, are
therefore zero\footnote{$^{**}$}{This is true for any two
times $t_1$ and $t_2$ despite
common misconceptions concerning the energy-time uncertainty principle}.

However, the quantum mechanics of closed systems does not only predict
probabilities for the outcomes of measurements of a subsystem.  We may
consider, if we wish, the probabilities of histories which describe
alternative values of the {\it total} value of a quantity commuting with
the {\it total} Hamiltonian $H$ for
the whole closed system at various moments of time.  Such total quantities
are unlikely to be accessible to experiment, but their conservation, or
lack of it, is still of theoretical interest, and it is this question which
is the subject of this paper.

The expression (1.5) for the probabilities of the histories of a closed system
would seem to allow non-zero probabilities for fluctuations
in a quantity commuting with the Hamiltonian if projections
that appear between
two projections associated with this conserved quantity
do not commute with it. However, in the
quantum mechanics of closed systems, probabilities are not predicted for
an {\it arbitrary} set of alternative histories.  They are predicted
only for those sets for which there
is negligible quantum mechanical interference between the individual
histories in the set [\Gria, \Omn , \GMH].   Such sets of histories
are said to decohere.
It would be inconsistent to assign probabilities to
sets of histories that did not decohere because
the correct probability sum rules would
not be obeyed.  Decoherence of histories implies the validity of the
probability sum rules so that decoherent sets of histories are consistent.

Conservation laws are obeyed by consistent sets of histories. In Section
\uppercase\expandafter{\romannumeral2}
we review an argument of Griffiths\footnote{$^{*}$}{The argument appears well
known to a number of people.  We learned it from R. Griffiths.} [\Grib]
that exact decoherence implies exact conservation of quantities that commute
with the Hamiltonian.
However, as we also show in Section
\uppercase\expandafter{\romannumeral2},
in a closed system of particles interacting by potentials, there are
severe limits on the exactly decohering sets of histories through
which the probabilities of fluctuations in a quantity commuting with the
Hamiltonian
could even be defined.  The only other alternatives
permitted in such histories are of the values of quantities that
effectively commute with the conserved quantity; i.e., commute when
acting on the initial condition of the closed system.  The probability
for any non-trivial evolution of such systems is zero.  That limitation
would
prohibit, for instance, consideration of a set of histories that
contained alternatives of the total energy as well as the alternatives
referring to position and momentum that would be needed to predict the
outcomes of our everyday observations.  (Recall that as observers we
are part of this closed system.)

However, a closed system of particles interacting via potentials is
not a realistic model of our universe.  The two most important
absolutely conserved quantities -- electric charge and mass -- are
coupled to long-range fields.  This fact has two
consequences: (1) It allows decoherent histories that describe possible
fluctuations in charge or mass together with other realistic,
everyday alternatives. (2) It ensures the exact decoherence of the
alternative values of these fluctuations, and
that the probability is zero for any non-vanishing value of a
fluctuation.  That is, total charge and total energy are exactly
conserved. The simplest case of electric charge is discussed in Section
\uppercase\expandafter{\romannumeral3}.
Section \uppercase\expandafter{\romannumeral4} discusses
the conservation of total energy.

\proclaim \uppercase\expandafter{\romannumeral2} Exact decoherence and Exact
Conservation.

In this section we review Griffiths' demonstration that the probabilities for
fluctuations in the values of quantities that exactly commute with the
Hamiltonian are exactly
zero for exactly decohering sets of alternative histories of a closed system.
We also show that, given a quantity $A$ which commutes with the
Hamiltonian, the only alternatives which can occur
in an exactly decohering set of histories describing possible
fluctuations of $A$
are values of operators which
effectively commute with $A$ when acting on the initial
condition of the system.

Let $A$ be any quantity satisfying
$$
[A,H] = 0
\eqno (2.1)
$$
including the Hamiltonian itself. Let $\{\Delta_\alpha\}, \alpha=1,2,\dots$
be an exhaustive
set of non-overlapping ranges of the eigenvalues of $A$, and let
$\{P^A_\alpha (t)\}$ be the set of Heisenberg picture projections onto
them.  The $\{P^A_\alpha(t)\}$ obey (1.4).
Consider a set of histories (consisting of alternatives at a sequence of
times) in which sets of projections onto  ranges of $A$ occur at two different
times $t_l$ and $t_m$.  The individual histories in such a set would be
represented by chains of projections operators
$$
C_\alpha = C^c_{\alpha_c}P^A_{\alpha_m}(t_m)
  C^b_{\alpha_b} P^A_{\alpha_l} (t_l) C^a_{\alpha_a}
\eqno (2.2)
$$
where the $\{C^a_{\alpha_a}\},\{C^b_{\alpha_b}\}, \{C^c_{\alpha_c}\}$
are the chains of projections representing alternatives before $t_l$,
between $t_l$ and $t_m$, and after $t_m$ respectively.  More generally the
$C^a_{\alpha_a},C^b_{\alpha_b},C^c_{\alpha_c}$ could be sums of chains
of projections corresponding to alternative histories defined by partitions
of the chains into exclusive classes, and they could be branch dependent in
the sense of [\HQC] without affecting the subsequent simple argument.

The decoherence functional whose off diagonal elements measure quantum
interference between parts of histories is
$$
D(\alpha',\alpha) = Tr ( C_{\alpha'}\rho C^\dagger_\alpha )
\eqno (2.3)
$$
where $\rho$ is the density matrix representing the initial condition of the
closed system.  When $Re(D)$ vanishes for $\alpha'\neq\alpha$ the set
of histories exactly (weakly) decoheres and the probabilities are given by
the diagonal elements, as summarized in the equation
$$
Re D(\alpha',\alpha) =\delta_{\alpha'\alpha} p(\alpha)\ .
\eqno(2.4)
$$
We can now proceed with Griffiths' argument.

Consider the probabilities
$p(\alpha_c,\alpha_m,\alpha_b,\alpha_l,\alpha_a)$ of the set of histories
represented by (2.2). Exact weak decoherence implies that these probabilities
are consistent.  That is, they must obey  the probability sum rules and in
particular
$$
\sum_{\alpha_a,\alpha_b,\alpha_c}
p(\alpha_c,\alpha_m,\alpha_b,\alpha_l,\alpha_a) = p(\alpha_m,\alpha_l)\
{}.
\eqno (2.5)
$$
The $p(\alpha_m,\alpha_l)$ are the probabilities for the set of histories
represented by the chain
$$
P^A_{\alpha_m}(t_m) P^A_{\alpha_l}(t_l) \ .
\eqno (2.6)
$$
But the individual operators in the chain are in fact {\it independent} of
$t$ because $A$ is conserved.  Specifically, the Heisenberg equations
of motion show that
$$
P^A_\alpha (t) = {\rm e}^{iHt/\hbar} P^A_\alpha (0) {\rm e}^{-iHt/\hbar}
=P_{\alpha}^A(0)
\eqno (2.7)
$$
because $A$ commutes with $H$. Thus
$$
P^A_{\alpha_m}(t_m) P^A_{\alpha_l}(t_l) = \delta_{\alpha_l\alpha_m}
P^A_{\alpha_l} (t_l)
\eqno (2.8)
$$
and the probabilities $p(\alpha_m,\alpha_l)$ which follow from (2.4)
vanish if  $\alpha_m\neq\alpha_l$.  Since the left hand side of (2.5)
is the sum of positive numbers, they must vanish individually.
We have
$$
p(\alpha_c,\alpha_m,\alpha_b,\alpha_l,\alpha_a) = 0 \ \ , \ \
\alpha_m\neq\alpha_l
\eqno (2.9)
$$
and the probability is zero for any non-vanishing fluctuation in the value of a
quantity that commutes with the Hamiltonian. Energy in particular is
conserved\footnote{$^*$}
{The argument for conservation depends only on the consistency of the
set of histories.  Although we introduced it by discussing weak
decoherence which implies consistency, the argument could proceed directly
from (2.5).}.

This satisfactory state of affairs is somewhat vitiated by the following
result
which shows that exact decoherence permits only alternatives values of
quantities that
effectively commute with the conserved quantity $A$ in between
times $t_l$ and $t_m$.

Suppose the set of histories represented by (2.2)
exactly decoheres.  Then every coarse graining of it must also exactly
decohere and in particular the set represented by
$$
C_{\alpha_m\alpha_b\alpha_l} = P^A_{\alpha_m} (t_m) C^b_{\alpha_b}
P^A_{\alpha_l} (t_l)\ .
\eqno (2.10)
$$
is exactly decoherent.
According to the result of Griffiths derived above, the probability of
a fluctuation in the value of $A$ is zero:
$$
p(\alpha_m,\alpha_b,\alpha_l) \equiv
Tr (C_{\alpha_m\alpha_b\alpha_l} \rho C^\dagger_{\alpha_m\alpha_b\alpha_l} )
=0 \ \ , \ \ \alpha_l\neq\alpha_m\ .
\eqno (2.11)
$$
Write the density matrix $\rho$ in the basis in which it is diagonal as
$$
\rho = \sum_\lambda |\lambda\rangle \pi_\lambda\langle \lambda |
\eqno (2.12)
$$
for positive probabilities $\pi_\lambda$.  In that basis, (2.11) reads
$$
\sum_{\lambda'\lambda} \pi_\lambda |\langle\lambda|
C_{\alpha_m\alpha_b\alpha_l}|\lambda'\rangle |^2= 0
\ \ , \ \ \alpha_l\neq\alpha_m
\eqno (2.13)
$$
so that
$$
C_{\alpha_m\alpha_b\alpha_l}|\lambda\rangle = 0,
\ \ {\rm if}
\ \ \alpha_l\neq\alpha_m
\ \ {\rm and\ }
\pi_\lambda \neq 0.
\eqno (2.15)
$$
Thus, $C_{\alpha_m\alpha_b\alpha_l}$ for $ \alpha_l\neq\alpha_m$ must vanish on
the subspace ${\cal S}_\rho$ of initial
states with non-vanishing probabilities in the
initial density matrix.  In particular
$$
C_{\alpha_m\alpha_b\alpha_l} \rho = 0 \ \ , \ \ \alpha_l\neq\alpha_m \ .
\eqno (2.16)
$$

The result (2.15) can be used to show that $C^b_{\alpha_b}$ must commute
with $A$ when acting on the subspace ${\cal S}_\rho$.   Suppose
$\{\Delta^m_{\alpha_m} \}$ and $\{\Delta^l_{\alpha_l} \}$ are sets
of ranges of
uniform, {\it infinitesimal} size $\Delta$ centered on eigenvalues
$a_{\alpha_m}$.
{}From (2.15) and (2.10) we can write
$$
(a_{\alpha_m}-a_{\alpha_l}) P^A_{\alpha_m}(t_m) C^b_{\alpha_b}
P^A_{\alpha_l}(t_l) |\lambda\rangle = 0, \ \ {\rm when}  \ \pi_\lambda\neq 0
\eqno (2.17)
$$
now holding for all values of $\alpha_m$ and $\alpha_l$.   In the  limit
of infinitesimal intervals $\Delta$, we have
$$
\sum_{\alpha_m} a_{\alpha_m} P^A_{\alpha_m} = A(t_m).
\eqno (2.18)
$$
Thus, by summing (2.17) over $\alpha_m$ and $\alpha_l$ we have
$$
[ A,C^b_{\alpha_b} ] |\lambda\rangle =0 , \ \ {\rm when} \ \ \pi_\lambda\neq 0
\eqno (2.19)
$$
and in particular
$$
[ A,C^b_{\alpha_b} ] \rho =0 \ .
\eqno (2.20)
$$
Thus the only permissible alternatives $C^b_{\alpha_b}$ in a set of
exactly decohering histories of the form (2.2) that test the
conservation of a quantity $A$ which commutes with the Hamiltonian
are alternatives of quantities that effectively commute with $A$
in the initial condition $\rho$.

The Hamiltonian corresponding to the total energy of course
commutes with itself.  The result (2.20) is sufficient to show that
histories of the form (2.2) that test conservation of energy can only
exhibit trivial dynamics. Consider
the case when $\{C^b_{\alpha_b}\}$ is a set of histories of alternative
ranges of values of a quantity $B$ at a time $t$ such that $t_l<t<t_m$.
Then,
$$
C^b_{\alpha_b} = P^B_\alpha (t)\ .
\eqno(2.21)
$$
{}From (2.20) we conclude
$$
[H, P^B_\alpha (t)]\rho = 0\ .
\eqno (2.22)
$$
Eq.(2.22) and the Heisenberg equation of motion
$$
P^B_{\alpha} (t) ={\rm e}^{iHt/\hbar} P^B_\alpha (0) {\rm e}^{-iHt/\hbar}
\eqno (2.23)
$$
are enough to show that for $t_l<t_1<t_2<t_m$
$$
P^B_{\alpha_2} (t_2) P^B_{\alpha_1}(t_1) \rho =0
            \ \ , \ \ \alpha_1\neq\alpha_2
\eqno (2.24)
$$
and this implies that there is {\it zero} probability of any change in the
value of $B$ for the histories in which fluctuations in the energy can be
defined.

To appreciate the strength of this result, imagine a model universe
consisting of a large box of
non-relativistic particles interacting by potentials.
Sets of histories describing just fluctuations in the total energy of
the model universe, say,
$$
C_\alpha = P^H_{\alpha_2} (t_2) P^H_{\alpha_1} (t_1)
\eqno (2.25)
$$
always exactly decohere.  That is because the $P^H_\alpha(t)$'s are
constant in time [{\it cf} (2.7)] so that the
$\{C_\alpha\}$ are a set of orthogonal {\it
projections}. The cyclic property of the trace in (2.3) shows that
histories consisting of orthogonal projections always decohere exactly.
Further, as a consequence of the orthogonality of the projections in
(2.25) for different ranges of the total energy, the $C_\alpha$
representing non-vanishing fluctuations in the total energy vanish
identically. The
probability of a fluctuation in the total energy is exactly zero.

However, the result (2.24) shows that it is not possible to fine grain the
histories (2.25) to include alternatives of any quantity other than the
total energy without
the values of that  quantity being constant in time
with probability one.  Thus,
were we part of such a system, we could not consider a set of histories that
describe the changes in our everyday lives and at the same time describe the
fluctuations in the total energy of the closed system.

The above discussion considered the conservation of precisely defined
values of quantities commuting with the Hamiltonian in exactly
decohering sets of histories in a model of particles interacting via
potentials.  It would be possible to consider the conservation of
quantities defined {\it im}precisely by a range of values but the results
would
depend on the details of the dynamics of the system studied.  It might
also be thought that it would be more realistic to consider approximately
decohering sets of histories and the approximate violations of
conservation laws that could be expected to occur.  However, more
important consequences can be derived
for interesting quantities like electric charge and mass by including
the long-range fields coupled to them that we have neglected until now.
We do this in the next two sections.

\proclaim  \uppercase\expandafter{\romannumeral3} Electric Charge.

The most firmly established examples of absolutely conserved
quantities  are electric charge  and total energy.  Both are
coupled to long-range fields and the conservation of each is
connected with a fundamental symmetry principle.  These fundamental
symmetry principles
limit the sets of decoherent histories
which can describe fluctuations in quantities that commute with the
Hamiltonian
and the values of the probabilities of these fluctuations.
The simplest case is electric charge which we discuss in this section.

The quantum theory of the electromagnetic field can be conveniently studied in
temporal,
$A_0(x)=0$, gauge where $A_\mu(x)$ are the components of the potential.
The states may be represented by vectors in the fermion Fock
space with components
that are functionals of the vector potential ${\vec A} ({\bf x})$.
Thus formulated
on a space which contains electromagnetic degrees of freedom beyond the true
physical
ones, the theory has a constraint represented by the operator
$$
{\cal C} (\Lambda) =  \int d^3x \Lambda ({\bf x}) [\nabla\cdot
{\vec E} ({\bf x}) -\rho({\bf x})]
\eqno (3.1)
$$
where $\rho({\bf x})$ is the charge density and ${\bf x}$ denotes the
three spatial coordinate of some
particular Lorentz frame.  The function $\Lambda ({\bf x})$
is the parameter of the gauge
transformation which is  generated by ${\cal C}(\Lambda)$ via
$$
\delta A_i ({\bf x}) = -{i \over {\hbar}} [{\cal C}(\Lambda), A_i({\bf x})] =
\partial_i \Lambda({\bf x})
\eqno (3.2)
$$
More generally, ${\cal C}(\Lambda)$ generates gauge
transformations for an arbitrary quasilocal (see, e.g. [\RH])
operator ${\cal O}$ that vanishes sufficiently fast outside some
bounded region of space (which we will call the effective
support) according to
$$
\delta {\cal O} = -{i \over {\hbar}} [{\cal C}(\Lambda), {\cal O}]\ .
\eqno (3.3)
$$
Physical, gauge invariant, quasilocal
operators must commute with the constraint
$$
 [{\cal C}(\Lambda), {\cal O}_{phys}] =0
\eqno (3.4)
$$
and physical, gauge invariant states are annihilated by the constraint.

As a particular case of  (3.4) we may take
$\Lambda=const.$  Then Gauss' law may be
applied to reduce the $\nabla\cdot {\vec E}$ term in (3.1) to a surface term
with spacelike separation from the effective support of
${\cal O}_{phys}$ yielding the result\footnote{$^{*}$}{This
derivation is naive form a rigorous
point of view.  For a
better one see [\SW].}
$$
 [Q, {\cal O}_{phys}] =0
\eqno (3.5)
$$
where
$$
Q=\int d^3x \rho(x)
\eqno(3.6)
$$
is the total charge operator.
Quasilocal physical quantities therefore commute with the total
charge\footnote{$^*$}{In
representing physical quantities by operators in this
way we are considering, as usual,
quantities defined at one moment of time.  For more
general spacetime alternatives see [\HarLH], Chap.VI.}.

An exhaustive set $\{ \Delta_\alpha \}, \alpha=1,2,\dots$ of ranges of a gauge
invariant quantity ${\cal O}$
define a set of alternatives for a closed system at a moment
of time.  These are represented by a
set of Heisenberg picture projection operators
$\{P^{\cal O}_\alpha (t)\}$.  Sets of histories
for the closed system may  be defined by
giving a series of such sets at a
sequence of times $t_1,\dots,t_n$.  The individual
histories correspond to particular sequences of  alternatives $(\alpha_1,\dots,
\alpha_n)$ and are represented by the
corresponding chains of  projectors as in (1.5).
More general examples of histories can be
obtained by partitioning such sequences
into classes $\{ c_\alpha \}$
represented by a set of class operators $\{ C_\alpha \}$
that are sums of the chains in the
class.  Thus, a gauge invariant set of
histories is generally represented by a set of class operators of the form
$$
C_\alpha = \sum_{(\alpha_1\dots\alpha_n)\in \alpha}
               P^{{\cal O}_n}_{\alpha_n} (t_n) \dots
P^{{\cal O}_1}_{\alpha_1} (t_1)\ .
\eqno (3.7)
$$

When electromagnetism is formulated in this way, with redundant as well as true
physical degrees of freedom,
the decoherence functional is not given by a formula like
(2.2) in which the Hilbert space is a space of wave functionals of the
vector
potential ${\vec A} ({\bf x})$.
Rather, it is given by that formula utilizing the Hilbert
space of functionals of the true physical
degrees of freedom; that is, just the transverse part $A^T({\bf x})$
of the vector potential.
The class operators (3.7) must first be reduced to class
operators $\{ C^T_\alpha \}$ on the
Hilbert space of physical degrees of freedom by integrating their
matrix elements over the redundant
(longitudinal) degrees of freedom with appropriate
gauge fixing conditions.   The details of this process are not important for
us\footnote{$^{**}$}{The construction
is standard, but for more details in the  present
notation see [\HarLH].};  it suffices to note
that the decoherence functional may be
written
$$
D(\alpha',\alpha) = Tr_T (C^T_{\alpha'} \rho {C^T_\alpha}^\dagger )
\eqno (3.8)
$$
where $\rho$ is a density matrix in the
Hilbert space of the matter degrees of freedom
and the true physical degrees of freedom of the electromagnetic field.

We next consider fine-graining a set of
histories (3.7) by including alternative values
of the total electric charge $Q$ at a
sequence of times $t'_k, k=1,\dots,m$.  We
consider, for simplicity the same set of  ranges
$\{\Delta_\beta\},\beta=1,2,3,\dots$ of
$Q$ at each of these times and let $\{P^Q_{\beta_k}
(t'_k) \}$ be the projections of the total charge operator onto them.  The
class
operators for such a finer grained set are
$$
C_{\alpha\beta} = \sum_{(\alpha\dots\alpha_n)\in \alpha}
                  	P^{{\cal O}_n}_{\alpha_n}(t_n)
\dots P^Q_{\beta_m}(t'_m) \dots
        \dots P^Q_{\beta_1}(t'_1) \dots	P^{{\cal O}_1}_{\alpha_1}(t_1)
\eqno (3.9)
$$
where the $P^Q_{\beta_k} (t'_k)$ have been
inserted at the positions dictated by
time ordering.

The projections $\{ P^Q_{\beta} (t')\}$ have
two important properties: First, they
commute with all gauge invariant quantities
as a consequence of (3.5), and therefore, in
particular
$$
[P^{{\cal O}_k}_{\alpha_k} (t_k),P^Q_{\beta_l} (t'_l)]=0\ .
\eqno (3.10)
$$
Second, they are conserved
$$
[H,P^Q_{\beta_l} (t'_l)] =0
\eqno (3.11)
$$
and therefore are independent of the times $t'_l$.
Eq.(3.10) means all
${P^Q}$'s may be commuted to the right or left in (3.9) and (3.11)
means that the class
operator is zero unless all the $\beta_l$ are the same
$$\eqalign{
C_{\alpha\beta} &=\delta_{\beta_n\beta_1}\dots
\delta_{\beta_2\beta_1} P^Q_{\beta_1} C_\alpha \cr
          & =\delta_{\beta_n\beta_1}\dots
\delta_{\beta_2\beta_1} C_\alpha  P^Q_{\beta_1}   \cr}
\eqno (3.12)
$$
The first of the relationships (3.12) shows that, for {\it any} set
of histories, the alternative values of the charge always decohere
{\it exactly}.
That is because, as a consequence of the cyclic property of the
trace, the decoherence functional $D(\alpha',\beta';\alpha,\beta)$
is always proportional to $\delta_{\beta'_1 \beta_1}$.
The $\delta$-functions in (3.12) thus ensure
that histories in which the total charge fluctuates have probability
zero. Total charge decoheres exactly and is exactly conserved.
Allowing approximate decoherence does not
permit non-zero probabilities for fluctuations in $Q$.

The restrictions derived in Section
\uppercase\expandafter{\romannumeral2} on histories that
include alternative values of the total charge
are still valid.  The alternatives in a decohering set of histories
must commute with the total charge.  However,
{\it all} quasilocal physical alternatives satisfy this condition
as a consequence of gauge
invariance.  It is therefore no restriction at all.

In general, if a set of histories $\{ C_\alpha\}$ decoheres,
then the finer-grained set (3.9) that includes
alternatives values of the charge does not necessarily decohere.  However, it
does in one interesting and natural case.
That is when the initial condition has a definite, fixed total
charge $q$.
$$
Q\rho = \rho Q = q\rho \ .
\eqno (3.13)
$$
Then when $C_{\alpha\beta}$, in the form of the second of
$(3.12)$, acts on $\rho$ there will be a non zero
result only for that interval $\Delta_\beta$,
which contains $q$.   $D(\alpha'\beta';\alpha\beta)$
is thus non-zero only when both $\beta'_1$ and $\beta_1$
have this value and is therefore
diagonal.  The finer-grained
set $\{C_{\alpha\beta}\}$ decoheres if the
set $\{C_\alpha\}$ does.  It follows that when the universe has a
definite value of the total charge\footnote{$^*$}{As it does for
instance in the ``no-boundary'' [\HH] initial condition
where the total charge is zero because the universe
is spatially closed.}, we may always fine-grain any
set of decoherent histories to ask
about the total charge without disturbing decoherence and receive
from the quantum mechanics of closed systems  the reassuring answer
that it is conserved with probability one.

Finally, note that all of our results follow directly from (3.5).
Any operator with the property satisfied by $Q$ in this equation is
said to be superselected (see, e.g. [\RH]).  In specific
restricted models, this occurs for quantities like baryon number,
lepton number, and non-Abelian charges as well as electric charge.
Thus, alternative values of superselected charges always decohere
exactly and are exactly conserved.  Furthermore, when the total
state has a definite value of such a charge, projections onto
its eigenvalues may be added to any set of histories without
affecting decoherence.

\proclaim  \uppercase\expandafter{\romannumeral4} Total Mass.

Energy universally couples to spacetime curvature which itself can carry energy
in the form of gravitational waves.  As a consequence, a realistic
classical discussion of the  total
energy of a closed system, which in relativity is the same thing as its
total mass,  must be carried out
in the context of general relativity and a
discussion of possible  quantum fluctuations in the total
energy in the context of quantum gravity.

There is no local definition of mass-energy in general
relativity because a general
spacetime does not exhibit a time-translation
symmetry.  Neither is it possible to define
the total mass of a spatially closed
cosmology except by assigning it the value zero in
which case its conservation  is trivial.
Conservation of energy becomes an interesting
issue in asymptotically flat spacetimes possessing
{\it asymptotic} time translation symmetries
enabling the {\it total} mass of the system to be
defined.

For asymptotically flat spacetimes, the
mass on a spacelike surface can be determined
from the asymptotic behavior of the
spatial metric on that surface.  Using coordinates
which asymptotically become rectangular
Minkowski coordinates at spatial infinity the
deviations from flat space must, at the very least, fall off as
$$
g_{\alpha\beta} = \eta_{\alpha\beta}
                  + {M_{\alpha\beta}(t,\theta,\phi) \over r} +
O \big( {1\over r^2}  \big)\  .
\eqno (4.1)
$$
Here, $\eta_{\alpha\beta}$ is the flat metric in rectangular
Minkowski coordinates,
$\eta_{\alpha\beta}= {\rm diag} (-1,1,1,1,)$,  $t=x^0$ and the polar
coordinates
$(r,\theta,\phi)$ are connected to the rectangular co\"ordinates
$(x^1,x^2,x^3)$ by the usual relations,
{\it e.g.} $x^1=r\sin\theta\cos\phi$.

Consider a spacelike surface that
asymptotically is a surface of constant $t$.  The ADM
total mass [10] is defined by evaluating the following
integral on a two-surface at large $r$
$$
M(t) = {1\over 16\pi} \lim_{r\rightarrow\infty}
 \int_t dS_j (\partial_k g_{kj} - \partial_j g_{kk} )\ .
\eqno (4.2)
$$
Here $\partial_i$ is the flat-space gradient and
we have followed the usual convention
of indicating a summation in asymptotic expressions
by repeated lower indices.   The asymptotic
behavior of the metric $(4.1)$ ensures that $M(t)$ is finite.

Whether total mass-energy is conserved in a quantum theory
of asymptotically flat spacetimes
depends on the probabilities of decoherent histories
that describe differing values of
$M(t)$ on different spacelike surfaces. There
are, of course, a variety of approaches to a quantum theory
of spacetime.  We shall
analyze the question in the sum-over-histories generalized quantum
theory of spacetime geometry.  A
generalized quantum theory is specified by three elements:
(1) The fine-grained
histories, which here are a class of four-dimensional metrics and matter
field configurations. The metrics $g_{\alpha\beta}(x)$ are
asymptotically flat at least in the sense of (4.1) but with possibly
more restrictive conditions to be discussed below and dwell
on a manifold with two spacelike boundaries $\sigma'$ and $\sigma''$
representing the ``endpoints'' of the history.
To keep the
notation manageable we shall indicate only a single matter field $\phi (x)$.
(2)  The
allowed coarse-grainings, which here are
{\it diffeomorphism invariant} partitions of the
fine-grained histories into exclusive classes $\{c_\alpha\},
\alpha=1,2,\cdots$
called coarse-grained histories.  (3) A
decoherence functional defining the measure of interference
between pairs of coarse-grained histories.
The precise details of the construction of
this decoherence functional will not be
important for us.  Its form is similar to (2.3) but with notions of $\rho$,
$Tr$, etc.  appropriate to gravity.  It is the form of the class operators
corresponding to coarse-grained histories that is important
for the present discussion of the conservation
of the total mass.   These class operators act on the space
of wave-functionals defined on the
space of three-metrics $h_{ij}({\bf x})$
and spatial matter field configurations
$\chi({\bf x})$ on a spacelike surface.
The matrix elements of the class operator
corresponding to  a diffeomorphism
invariant class  $c_\alpha$ of asymptotically flat
four geometries and field configurations are defined by the
sum-over-fine-grained-histories:
$$
\langle h_{ij}'',\chi'' || C_{\alpha}|| h_{ij}',\chi' \rangle =
\int_{[(h',\chi'),c_\alpha,(h'',\chi'')]} \delta g \delta\phi\ {\rm exp}
\{iS[g(x),\phi(x)]/\hbar\}\  .
\eqno (4.3)
$$
Here, $h'_{ij}({\bf x})$ and $\chi'({\bf x})$ are the
induced metrics and matter field configurations
on the boundary $\sigma'$.  There are similar definitions on $\sigma''$.
$S[g,\phi]$ is the
action for geometry coupled to matter fields.  The sum is over four-metrics
$g_{\alpha\beta}(x)$ and
four dimensional field configurations $\phi(x)$ which are in
the diffeomorphism invariant class $c_\alpha$ and match
the prescribed conditions on
$\sigma'$ and $\sigma''$.  Of course,
the expression (4.3) is only formal and must be
augmented by gauge fixing machinery, regularization procedures, {\it etc}
to make sense, but its form
will be sufficient for the level of
argument we are able to
give. Further details can be found in, for example [\HarLH].

With these preliminaries in hand we may return to the issues
of the conservation of total
ADM mass at  spatial infinity and whether histories that define
fluctuations in the total mass are limited to trivial dynamics as they were in
the simple
model of Section II which neglected gravitation.  To calculate the probability
of a fluctuation in the
mass, we must consider partitions of the set of fine-grained histories
into classes by ranges
of the value of the total
mass $M(\sigma)$ on at least two different spacelike surfaces
$\sigma_1$ and $\sigma_2$,
in addition to whatever other alternatives define the classes under
consideration. Such histories
are the analog of those represented by (2.2) when $A$ is the total
energy, $H$, in the
non-gravitational case.   When such sets decohere, the issue of
conservation of ADM
mass is then the question of
whether the probability is zero for those
with $M(\sigma_1) \neq M(\sigma_2)$.
For this case, the arguments of Section III are not satisfactory as
quasilocal diffeomorphism invariant operators are
difficult to construct -- in fact, because there are no
{\it local} diffeomorphism invariant operators
for gravity, strict use of the definition
in [\RH] shows that there are {\it no} quasilocal invariant operators
at all.
To find a more satisfying
argument we must look more closely at what is
meant by ``asymptotically flat''.

Penrose's notion of conformal completion
[\Pena,\Penb] gives a standard definition of a
spacetime which is asymptotically flat\footnote{$^*$}{For a
lucid review see
Ashtekar[\Ash]}.  A consequence of this definition is that
asymptotically  flat metrics have a more restricted asymptotic behavior
than that given by (4.1).
In particular the Riemann tensor must decay at large $r$ as
$$
R_{\alpha\beta\gamma\delta} = O \big (1 / r^3 \big).
\eqno (4.4)
$$
It is not difficult to show that, in any $3+1$ decomposition of spacetime
into space and time, this implies
$$
\dot M_{ij} =0
\eqno (4.5)
$$
where a dot  denotes a time derivative and Roman indices range over
spatial directions.
This means that the ADM mass, as defined by (4.2)
is constant in time.   The conservation of ADM mass in
this context does not follow  from
the equation of motion, but from the definition of an
asymptotically flat spacetime.  Of course, the
asymptotically flat context would be uninteresting except that solutions
of this form {\it do} exist.  The finite propagation velocity of
gravitational radiation ensures that any solution with suitably
localized initial data will be asymptotically flat.

The Penrose diagram for the conformally completed asymptotically
flat spacetime makes the reason for this ``conservation"
intuitively clear.  Spacelike infinity is a single two-sphere where all
spacelike surfaces terminate.  A common value of the ADM mass is
therefore shared by all.

Were we to use conformal completion to define the
asymptotically flat metrics which enter into
the sum-over-histories (4.1) the question of conservation of total mass
would be trivial.
Only geometries with constant total mass contribute to the sum,
therefore partitions into
classes with different masses on different spacelike surfaces would be vacuous.

However, while the conservation of total mass at spatial infinity
is trivial, the dynamics permitted in histories that define this
conservation is not.  In the model without long-range fields discussed
in Section II, only alternatives of quantities that effectively
commuted with the total energy were permitted in exactly decohering
sets of histories which also described fluctuations in the total
energy.  However, in the presence of the gravitational field, the
analog of (2.17) which led to that result is trivially satisfied
for {\it any} diffeomorphism invariant partition of the fine-grained
histories, regardless of whether it is associated with projections onto
eigenvalues of quasi-local operators.  That is because there are no
fine-grained histories at all with fluctuations in the total mass.
We therefore expect that in generalized quantum theory we are permitted
arbitrary sets of physical histories that also
describe fluctuations in the total
mass.  If
any set of alternatives decoheres, we may always consider the finer
graining which in addition describes fluctuations in the total mass.
If that finer graining continues to decohere, the alternatives
referring to the total mass decohere exactly.  Total mass, or total energy
which is the same thing, is
conserved with probability one.

The above discussion was carried out using the conformal completion
definition of asymptotic flatness.
However, from the perspective of quantum gravity it appears
more natural to define
asymptotic flatness from a property of the {\it action} rather than from
a  notion of conformal
completion.  We now show that if the sum-over-histories in
(4.3) is restricted to a class
of metrics with the fall-off (4.1)
that (1) have finite action and (2) are invariant under
diffeomorphisms, then the ADM mass is conserved.  To understand
this it is sufficient to look at the action for pure gravity.

The action for gravity on a domain of spacetime ${\cal M}$ is
$$
 (16\pi G)S_E[g] = \int_{\cal M} d^4x \sqrt{-g} R
               +  2\int_{\partial{\cal M}} d^3x \sqrt{h} K
\eqno (4.6)
$$
where $R$ is the scalar curvature and $K$
is the extrinsic curvature scalar of the boundary of ${\cal M}$.  In order to
discuss the properties of metrics at
spatial infinity, it is useful to consider a
standard $3+1$ decomposition of the metric
$$
ds^2= -N^2 dt^2 + h_{ij} (dx^i +N^idt)(dx^j+N^jdt).
\eqno (4.7)
$$
which need only hold near spatial infinity for our purposes.
Consider the action for a region of
spacetime lying between two spacelike surfaces of constant $t$ and
bounded by a timelike surface $\partial{\cal M}_s$ near infinity. This
may be written as
$$\eqalign{
(16\pi G)S_E[g] =& \int dt \int d^3x Nh^{1/2} [ K_{ij}K^{ij} - K^2 + ^3R]   \cr
               & - \int _{\partial{\cal M}_s} d\Sigma_i (-2 KN^i + 2D^iN)
                +  2\int_{\partial{\cal M}_t} d^3x \sqrt{\tilde h}
(\tilde K-\tilde K_0) \cr}
\eqno (4.8)
$$
where $K_{ij}$ is the extrinsic curvature
of surfaces of constant $t$,
$$
K_{ij} = {1\over 2N} \Big [ -\dot h_{ij} + D_{(i} N_{j)} \Big ]\ ,
\eqno (4.9)
$$
$D_i$ being the derivative in the surface, $\tilde K_{ij}$ is the
extrinsic curvature of $\partial{\cal M}_t$,  and the tilde
indicates that a quantity is to be
evaluated for a timelike surface.

The general form of the metric (4.1) is not sufficient to
ensure the convergence of the
action (4.8).  Eq.(4.1) implies that the asymptotic behavior of
$^3 R$ is $O(1/r^3)$ but
the asymptotic behavior of $K_{ij}$ is
$$
K_{ij} = -{1\over 2r} \dot M_{ij} + O ({1\over r^2})\ .
\eqno (4.10)
$$
If we evaluate the volume term in the action (4.8)
out to a large radius $r_l$,
the coefficient of the leading term  as $r_l\rightarrow \infty$ is
$$
r^2_l \int dt \int d\Omega [ (\dot h_{ij})^2 - (\dot h_{kk})^2 ]
     = \int dt \int d\Omega [ (\dot M_{ij})^2 - (\dot M_{kk})^2 ]
\eqno (4.11)
$$
where $d\Omega$ is an element of solid angle at  infinity.
Thus metrics in the class
(4.1) must be further restricted so that the right hand side of (4.11)
vanishes in order to
ensure finite action.  If we assume that this
should hold for any choice of the time
interval between the boundary surfaces we must have
$$
r^2_l  \int d\Omega [ (\dot h_{ij})^2 - (\dot h_{kk})^2 ]
     =  \int d\Omega [ (\dot M_{ij})^2 - (\dot M_{kk})^2 ]
     = 0\ .
\eqno (4.12)
$$
This is not enough to show that the ADM mass is
constant in time, but when coupled with the
requirements of diffeomorphism invariance it will be.

The asymptotic behavior of (4.1) refers to a particular decomposition
of the spacetime
into space and time, and the condition (4.12) ensures that there is  no linear
divergence of the action when evaluated between two constant time
surfaces in that
decomposition.  However, since the class operators
(4.3) are to be defined by integrals
over diffeomorphism invariant partitions, the notion of
asymptotic flatness and of
finiteness of the action must be independent of the
$3+1$ decomposition.  In particular
it must be invariant under diffeomorphisms which Lorentz transform
the asymptotic slices.
This leads to stronger conditions than (4.12) as we shall show.

Consider infinitesimal diffeomorphisms (gauge-transformations)
$$
g_{\alpha\beta}(x) \rightarrow g_{\alpha\beta} +
\nabla_{(\alpha}\xi_{\beta)}(x)\ ,
\eqno (4.13)
$$
and in particular those which asymptotically correspond to Lorentz boosts
$$\eqalign{
\xi^i &\approx v^it + d^i(t,\theta,\phi) + O (1/r)\ , \cr
\xi^0 &\approx v_i x^i + O(1)\ .   \cr}
\eqno (4.14)
$$
Lorentz boosts preserve the asymptotic behavior (4.1).  The
supertranslations $d^i$, however, must be independent of time to
preserve (4.1), as
substitution into (4.7) will show.  The covariant components of $\xi_i(x)$
relevant for
the transformation of the $h_{ij}(x)$ are thus
$$
\xi_i \approx v_i t + d_i (\theta,\phi) +  N_i (x) (v_jx^j)  + O (1/r)
\eqno (4.15)
$$
displaying explicitly the $O(r)$ and $O(1)$ terms.   Since $N_i(x)\approx
s_i(t,\theta,\phi)/r$ the third term is of $O(1)$.
The spatial part of the metric
important for the asymptotic form of the metric thus transforms as
$$
h_{ij}\rightarrow \delta_{ij} + {M_{ij}\over r}  + \partial_{(i}\xi_{j)}
+ O({1\over r^2})
\eqno (4.16)
$$
with $\xi_j(x)$ of the form (4.15).

Diffeomorphism invariance requires that the condition (4.11)
be enforced for $h_{ij}(x)$ of the
form (4.16) with $\xi_i(x)$ given by (4.15).
The first term in (4.15) does not change
the spatial metric.  In determining the effect of the
rest of (4.15) on $h_{ij}(x)$
the boost parameter $v_i$ is arbitrary.
But it is also important to
note that the $s_i(t,\theta,\phi)$ determining $N_i(x)$
is arbitrary;  $s_i$ merely defines
how the spacetime is sliced internally,
consistent with a given asymptotic slicing.  We
must therefore enforce the condition (4.12)
for $h_{ij}(x)$ of the form (4.16) with
arbitrary $\xi_i(x)$ of $O(1)$.
One further invariance should be enforced.  In (4.11)
we evaluated the action inside spheres of constant $r_l$
and considered the limit
$r_l\rightarrow \infty$.   The same results should hold for arbitrary
shaped surfaces
$r_l=r_l(\theta,\phi)= Rf(\theta,\phi)$ as $R\rightarrow \infty$.
To first order in $\xi_i(x)$, the gauge transformed
condition (4.12) becomes the condition that the linear
divergence in $R$ of
$$
\int^{r_l(\theta,\phi)} r^2 dr \int d\Omega [ u_{ij}(x) \partial_i\xi_j(x)]
\eqno (4.17)
$$
vanish.  Here we have abbreviated
$$
u_{ij}(x) = \dot h_{ij}(x) -\delta_{ij}\dot h_{kk}(x)
\eqno (4.18)
$$
and assumed, as discussed above, that the $O(1)$ part of
$\xi_i(x)$ is an arbitrary function of $(t,\theta,\phi)$.
Integrating (4.17) by parts and retaining only the leading terms in
large $R$ in the resulting
condition following from the arbitrary form of $\xi_i(x)$ we have
$$
R[-R f(\theta,\phi)\partial_i u_{ij}(x) + n_i(\theta,\phi) u_{ij} (x) ] =0
\eqno (4.19)
$$
for large $R$ where $n_i$ is the normal to the bounding surface proportional to
$\partial_if$.  Since $f$ is arbitrary (4.19) can be
satisfied only if both terms vanish and, in particular, if $Ru_{ij}(x)=0$.
That implies
$$
\dot M_{ij} = \delta_{ij} \dot M_{kk}\ .
\eqno (4.20)
$$
This condition together with (4.12) is enough to guarantee
$$
\dot M_{ij}(t,\theta, \phi) = 0\ .
\eqno (4.21)
$$

The result (4.21) is enough to guarantee that the
Riemann tensor falls off as $O(1/r^3)$
as in (4.4) and ensure the conservation of the ADM mass.
To see that, simply note that
from (4.2), the ADM mass is determined by the coefficients
$M_{ij}$ all of whose time
derivatives vanish.

Thus restricting attention to metrics with the minimal
asymptotic behavior (4.1) for
which the action is finite and diffeomorphism invariant means
restricting to metrics for which
the Riemann tensor falls off as $O(1/r^3)$
and for which the ADM  mass is constant.   It
is an interesting question whether the above rather
clumsy argument could be pushed
further and whether there is full
equivalence between asymptotic flatness defined by
finiteness and diffeomorphism invariance of the gravitational action and
asymptotic flatness
defined by conformal completion.  It would be of special
interest to investigate and
compare the conditions necessary to define total angular
momentum at spatial infinity,
a quantity that we have not touched upon.

Thus, whether asymptotic flatness is defined by conformal completion
or by the behavior (4.1), finiteness of the action, and covariance, the
result is the same for the ADM mass.  It is conserved in each history.
The set of fine-grained histories includes histories with differing
values of the ADM mass but within each history it does not vary.
Arbitrary diffeomorphism invariant alternatives may therefore be considered in
addition to those necessary to describe the
conservation of total mass
of a closed system and, if all alternatives
decohere, total mass, which is the same as the total energy, is conserved.

\proclaim Acknowledgments.

We would like to thank the Aspen Center for Physics where this work
was initiated and also the Isaac Newton Institute for Mathematical Sciences.
We would also like to thank Bob Griffiths, Jonathan Halliwell, Karel
Kucha\v{r}, Roger Penrose, John Stewart and Wojciech Zurek for useful
conversations.  The work of J.H. was supported in part by NSF grant
PHY90-08502.  R.L. thanks Los Alamos National Laboratory for support.
D.M. was supported in part by NSF grant PHY93-96246 and funds provided
by the Pennsylvania State University.

\proclaim References.

\item{\Gria .} R. Griffiths, J. Stat. Phys. {\bf 39}, 219 (1984).

\item{\Omn .} R. Omn\`es, J. Stat. Phys. {\bf 53} 893, (1988); {\bf 53},
957 (1988); {\bf 53}, 993 (1988); {\bf 57}, 357 (1989); Rev. Mod. Phys.
{\bf 64}, 339 (1992).

\item{\GMH .} M. Gell-Mann and J.B. Hartle in {\it
Complexity, Entropy and the Physics of Information},
SFI Studies in the Sciences of Complexity, Vol. VIII,
ed by W. Zurek  (Addison-Wesley, Reading, 1990).

\item{\HQC .} J. B. Hartle, in {\it Quantum Cosmology and Baby
Universes}, Proceedings of the 1989 Jerusalem Winter School for
Theoretical Physics, edited by S. Coleman, J. B. Hartle, T. Piran, and
S. Weinberg (World Scientific, Singapore, 1991).

\item{\Grib .} R. Griffiths, Private communication.

\item{\RH .} R. Haag, {\it Local Quantum Physics}  (Springer-Verlag, New
York, 1992).

\item{\SW .}  F. Strocchi and  A. S. Wightman,
J. Math. Phys. {\bf 15}, 2198 (1974).

\item{\HarLH .} J. B. Hartle in {\it Gravitation and Quantization},
Proceedings of the 1992 Les Houches Summer School, ed. by B. Julia
and J. Zinn-Justin, Les Houches Summer School Proceedings Vol LVII
(North Holland, Amsterdam, 1995),  gr-qc/9304006.

\item{\HH .} J. B. Hartle and S. W. Hawking, Phys. Rev. D {\bf 28}, 2960
(1983).

\item{\ADM .} R. Arnowitt, S. Deser, and C.W. Misner in {\sl
Gravitation:
An Introduction to Current Research}, ed. by L. Witten (Wiley,
New York, 1962) and references to earlier work therein.

\item{\Pena .}  R. Penrose, Phys. Rev. Lett. {\bf 10}, 66 (1963).

\item{\Penb .}  R. Penrose,  Proc. Roy. Soc. London, {\bf A284}, 159 (1965).

\item{\Ash .}  A. Ashtekar, in General Relativity and Gravitation,
Vol. II, ed. by A.Held (Plenum Press, N.Y., 1980)  p.37ff.

\item{\GMHB .} M. Gell-Mann and J. B. Hartle, Phys. Rev. D {\bf 47},
3345 (1993), gr-qc/9404013.

\bye